\documentclass[a4paper,twocolumn,11pt,accepted=2023-08-10]{quantumarticle}

\pdfoutput=1
\usepackage[utf8]{inputenc}
\usepackage[english]{babel}
\usepackage[T1]{fontenc}
\usepackage{amsmath}
\usepackage{tikz}
\usepackage{graphicx}
\usepackage{epstopdf}
\usepackage{lipsum}

\usepackage{mathrsfs}
\usepackage[retainorgcmds]{IEEEtrantools}

\usepackage{physics}
\usepackage{amssymb}
\usepackage[amssymb]{SIunits}
\usepackage{ulem}
\usepackage{color}

\usepackage{bm}
\usepackage{mathrsfs}
\usepackage[retainorgcmds]{IEEEtrantools}
\usepackage{amsmath}
\usepackage{amsfonts}
\usepackage{times,txfonts}
\usepackage{nicefrac}
\usepackage[colorlinks=true,linkcolor=magenta,urlcolor=magenta,citecolor=magenta,pdfusetitle]{hyperref}

\newcommand{\up}{\uparrow}
\newcommand{\dwn}{\downarrow}
\newcommand{\U}{\Uparrow}
\newcommand{\D}{\Downarrow}

\usepackage[backend=bibtex, 
            isbn=false, 
            eprint=false, 
            url=false,
            sorting = none]{biblatex}
\DefineBibliographyStrings{english}{bibliography = {References}}
            
\begin{document}

\title{Energy-efficient quantum non-demolition measurement with a spin-photon interface}
\author{Maria Maffei}
\affiliation{Dipartimento di Fisica, Università di Bari, I-70126 Bari, Italy}

\author{Bruno O. Goes}
\affiliation{Université Grenoble Alpes, CNRS, Grenoble INP, Institut Néel, 38000 Grenoble, France}

\author{Stephen C. Wein}
\affiliation{Université Grenoble Alpes, CNRS, Grenoble INP, Institut Néel, 38000 Grenoble, France}
\affiliation{Quandela SAS, 10 Boulevard Thomas Gobert, 91120 Palaiseau, France}

\author{Andrew N. Jordan}
\affiliation{ Institute for Quantum Studies, Chapman University, 1 University Drive, Orange, CA 92866, USA}
\affiliation{Department of Physics and Astronomy, University of Rochester, Rochester, New York 14627, USA}

\author{Loïc Lanco}
\affiliation{Université Paris Cité, Centre for Nanoscience and Nanotechnology (C2N), F-91120 Palaiseau, France}

\author{Alexia Auffèves}
\affiliation{MajuLab, CNRS–UCA-SU-NUS-NTU International Joint Research Laboratory}
\affiliation{Centre for Quantum Technologies, National University of Singapore, 117543 Singapore, Singapore}

\begin{abstract}
Spin-photon interfaces (SPIs) are key devices of quantum technologies, aimed at coherently transferring quantum information between spin qubits and propagating pulses of polarized light. We study the potential of a SPI for quantum non demolition (QND) measurements of a spin state. After being initialized and scattered by the SPI, the state of a light pulse depends on the spin state. It thus plays the role of a pointer state, information being encoded in the light's temporal and polarization degrees of freedom. Building on the fully Hamiltonian resolution of the spin-light dynamics, we show that quantum superpositions of zero and single photon states outperform coherent pulses of light, producing pointer states which are more distinguishable with the same photon budget. The energetic advantage provided by quantum pulses over coherent ones is maintained when information on the spin state is extracted at the classical level by performing projective measurements on the light pulses. The proposed schemes are robust against imperfections in state of the art semi-conducting devices. 
\end{abstract}

\maketitle

%
%

\section{Introduction}
A spin-photon interface (SPI) is a device, whose purpose is to coherently transfer information between a spin, which plays the role of a storage qubit, and a propagating pulse of polarized light, i.e. a flying qubit. Experimental implementations range from atomic physics \cite{Rempe2007}, ion traps \cite{blatt2012}, to semi-conductor devices \cite{gao_observation_2012,javadi_spinphoton_2018}. SPIs are key components to implement a variety of functionalities, from quantum memories and quantum repeaters \cite{kimble_quantum_2008}, to photon-photon gates \cite{hu_giant_2008,bonato_cnot_2010} and cluster states \cite{schwartz_deterministic_2016,Fioretto2022,cogan2021deterministic}, which are essential for light-based quantum technologies. 

Most SPI functionalities rely on the capacity to extract reliable information on the spin state by measuring the light state. This encompasses the ability to coherently map information from the spin to the light, and to perform well-chosen projective measurements on light pulses. These steps are constitutive of a von Neumann measurement scheme \cite{neumann_mathematical_2018} where light plays the role of a quantum meter that first maps the spin state in the readout basis (pre-measurement), before being collapsed.

\begin{figure}
\includegraphics[width=0.5\textwidth]{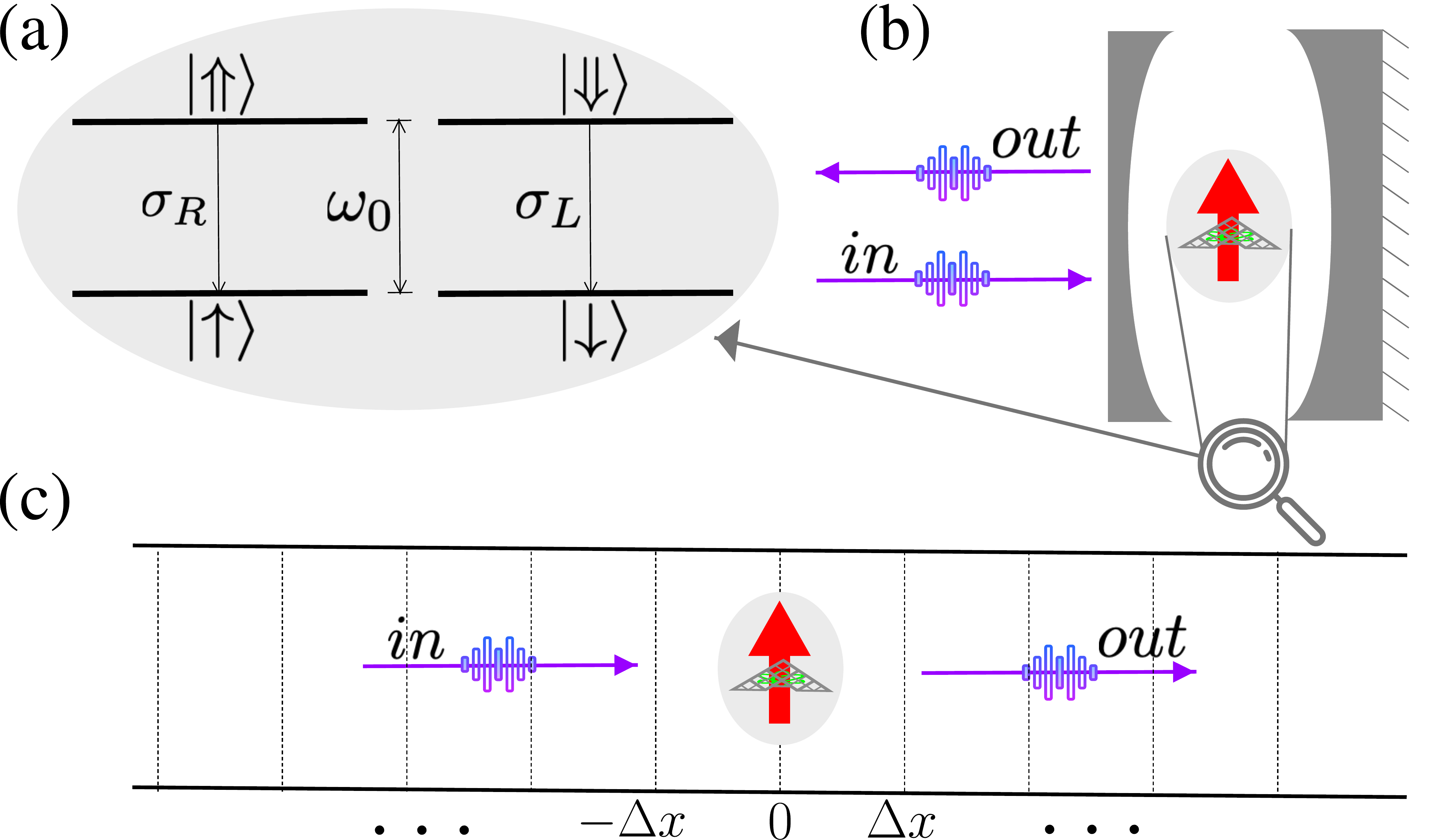}
\caption{Scheme of the spin-photon interface. a) Energetic structure of the quantum emitter. The system is a 4-level system hosting two degenerate transitions respectively coupled to left (right) circularly polarized light. b) The interface consists of a quantum emitter with a spin degree of freedom (red arrow in the gray shaded area) and coupled with a 1D waveguide. c) Unfolded wave-guide. The SPI is positioned at $x=0$ and interacts sequentially with the input ancillas. }
\label{Fig1-1}
\end{figure}

Here we analyze the performance of a SPI to achive quantum non demolition (QND) measurements of the spin state. We first consider the pre-measurement step and then the full measurement scheme. We choose the quantum and classical Bhattacharyya coefficients \cite{fuchs_cryptographic_1999} as respective figures of merit. In the spirit of quantum metrology \cite{maccone_science_2004}, we optimize the readout performance as a function of the state of the light probe. In particular, we compare the use of classical and quantum resources, by computing the spin-light dynamics for coherent pulses and superpositions of zero and single photon pulses respectively. We find that quantum resources reach better performances than classical ones with the same photon budget. Such energetic advantage of quantum nature is usually chased in optical quantum metrology, which operates at low-intensity probes \cite{Pan2023}. It could also become a key incentive to deploy quantum technologies at large scales \cite{QEI}.   

Importantly, our analysis relies on a complete Hamiltonian model of the spin-light dynamics, based on a recent extension of the so-called collision model \cite{Ciccarello_review_2021, ciccarello_collision_2017, maffei_closed-system_2022}. Spin-light entangled states are computed at any time, taking into account all the light temporal modes and the pulse deformations induced by the coupling to the spin. This represents an important progress with respect to state of the art models, where single monochromatic light modes is usually considered. It opens new opportunities to control and optimize non-ideal behaviors in light-based quantum devices, where light pulses must have a finite duration.   

Our article is organized as follows. We first introduce the SPI under study, and recall basic but essential results related to light scattering by a single qubit. We then analyze the performance of the SPI as a measuring device for the spin state. We find a quantum advantage, both at the levels of the pre-measurement and full measurement scheme. We discuss the origin of such a quantum advantage in our study and quantum metrology. We finally provide a feasibility study of the proposed experiments and show that the quantum advantage can be observed with state of the art devices.

\section{System and model}
The SPI features a 4-level system (4LS) (see Fig.\ref{Fig1-1}(a)). It is composed of two ground states, $\left\{\ket{\uparrow},\ket{\downarrow}\right\}$, with spin projections respectively $\pm \hbar/2$, and zero energy; and two excited states, $\left\{\ket{\Uparrow},\ket{\Downarrow}\right\}$, with spin projections respectively $\pm 3\hbar/2$ and energy $\hbar\omega_0$. This level scheme is typical of an electronic spin trapped in a quantum dot~\cite{lindner_proposal_2009}. It gives rise to two degenerate transitions respectively coupled to circularly polarized electromagnetic fields. The 4LS bare Hamiltonian reads:
\begin{align}\label{eq_spinbare}
H_{\text{4LS}}=\hbar\omega_0\sum_{j=\text{R,L}}\sigma_{j}^{\dagger}\sigma_{j},
\end{align}
where we defined the lowering operators $\sigma_{\text{L}}=\ket{\downarrow}\bra{\Downarrow}$ and $\sigma_{\text{R}}=\ket{\uparrow}\bra{\Uparrow}$. Due to the conservation of the angular momentum, the transition $\ket{\downarrow}\rightarrow\ket{\Downarrow}\left(\text{resp.} \ket{\uparrow}\rightarrow\ket{\Uparrow}\right)$ is coupled to left (right) circularly polarized light pulses.

The 4LS is positioned at the position $x=0$ of a waveguide (WG), where light can only propagate in one direction (see Fig.~\ref{Fig1-1}(c)). The WG hosts two reservoirs of circularly polarized modes of frequencies denoted $\omega_k$ with lowering operators $a_{j,k}$, where $j \in \lbrace \text{R},\text{L}\rbrace$ stands for right- and left-circular polarization. The field dispersion relation reads $k=\omega_k v^{-1}$ where $v$ is the field group velocity and $k \geq 0$ its wave vector. We only kept the positive values of $k$, which captures the unidirectionality of the field propagation. Hence, the WG bare Hamiltonian reads:
\begin{align}\label{eq_field_bare}
H_{f\text{pol}}=\hbar\sum_{j=\text{R,L}}\sum_{k=0}^{\infty}\omega_{k}a_{j,k}^{\dagger}a_{j,k}.
\end{align}
Let us notice that this situation does not correspond to the so-called chiral waveguides, where the direction of propagation depends on the light polarization~\cite{lodahl2017chiral} - here all polarizations propagate in the same direction.

This boils down to coupling the emitter to a single input-output port, as it was primarily considered in the seminal paper introducing the input-output formalism \cite{gardiner_input_1985}, and since then in various theoretical works, e.g. \cite{kojima_efficiencies_2004, gross_qubit_2018, Fan_inout, Fischer,Molmer2019,maffei_probing_2021}. The SPI coupling Hamiltonian is then:
\begin{align}\label{eq_spi_coupling}
H_{\text{SPI}}=i g\hbar\sum_{j=\text{R,L}} \sum_{k=0}^{\infty} \left[\sigma_j^\dagger a_{j,k}-a_{j,k}^{\dagger} \sigma_j \right],
\end{align}
where we implicitly assumed that the light-matter coupling $g$ is the same for both polarizations and uniform in frequency. This assumption is valid when the coupling is weak enough that only frequency modes close to $\omega_0$ play a role (quasi-monochromatic approximation)~\cite{gardiner_input_1985,gross_qubit_2018}. In this regime, the rotating wave approximation is allowed~\cite{loudon_quantum_2000}.

Coupling an emitter to a unidirectional light fields is challenging and most experimental situations correspond to a quantum emitter coupled to multiple input-output ports. However, there are some cases where the unidirectional model is accurate as argued in \cite{Hofmann_2003} - in particular, when the emitter is weakly coupled to an asymmetric, directional cavity (see Fig.~\ref{Fig1-1}(b)). This is the case for a quantum emitter embedded in an asymmetric Fabry-Perot cavity, itself coupled to an optical fiber \cite{hunger_fiber_2010} , or to a free-space gaussian beam with proper impedance matching \cite{hilaire_accurate_2018}. Adiabatic elimination of the cavity \cite{carmichael_statistical_2008} yields an effective, unidirectional atom justifying a WG-QED treatment.

We consider the scattering of a light pulse by the 4LS. All along the paper, we shall refer to the input (resp. output) field as the initial light state at $t=0$ (resp. to the scattered light state at $t\rightarrow +\infty$). Let us consider a L-polarized (resp. R) input pulse $\ket{\psi_{\text{L}}}$ (resp. $\ket{\psi_{\text{R}}}$). Initial states of the kind $\ket{\up}\otimes\ket{\psi_{\text{L}}}$ and $\ket{\dwn}\otimes\ket{\psi_{\text{R}}}$ are preserved by the scattering process. Conversely, $\ket{\psi_{\text{R}}}$ (resp. $\ket{\psi_{\text{L}}}$) interacts with the 4LS if the spin is in the state $\ket{\up}$ (resp. $\ket{\dwn}$), giving rise to a dynamics equivalent to that of a spinless 2-level system (2LS) interacting with a non-polarized propagating field. This case has been solved analytically for an input pulse being either a coherent state~\cite{Fischer,maffei_closed-system_2022} or a superposition of zero and one photon~\cite{Fan_inout,maffei_closed-system_2022}, providing the exact light-matter states at any time. Below we recall the solutions obtained for such a spinless 2LS, that we will use later on to derive those of the 4LS featuring the SPI.  

\section{Scattering by a 2LS}
Following Ref. \cite{maffei_closed-system_2022}, we consider a 2LS
positioned at the point $x=0$ of some unidirectional WG. The states of the 2LS are the ground and excited states
$\{\vert g\rangle,\vert e\rangle\}$. The field is assumed to propagate
from left to right with velocity $v$. The annihilation operator $a_{k}$ destroys
a photon with positive wave vector $k$ and frequency $\omega_{k}=v k$. The total Hamiltonian reads
\begin{equation}
H=H_{\text{2LS}}+H_{f}+H_{I}
\end{equation}
with 
\begin{align}
H_{\text{2LS}}&=\hbar\omega_{0}\sigma^{\dagger}\sigma,~
H_{f}=\hbar\sum_{k}\omega_{k}a_{k}^{\dagger}a_{k},\\ \nonumber
H_{I}&=ig \hbar \sum_{k}\left[\sigma^{\dagger}a_{k}-a_{k}^{\dagger}\sigma\right].
\end{align}
Let us notice that we extended the lower limit of the summation over $k$ to $-\infty$ in order to define the Fourier transform of $a_k$~\cite{Fan_inout}. The dynamics is solved in the interaction picture with respect to $H_{f}+H_{{\text{2LS}}}$, yielding the interaction Hamiltonian,
\begin{equation}
H_{I}(t)=i\hbar\sqrt{\gamma}\left[\sigma^{\dagger}(t)b(t,0)-b^{\dagger}(t,0)\sigma(t)\right].\label{eq:Int-pic-Hamiltonian}
\end{equation}
We defined the emitter's dipole in the interaction picture as $\sigma(t)=e^{-i\omega_{0}t}\sigma$, and the annihilation operators destroying excitations located in the position $x$ at the time $t$~\cite{gross_qubit_2018,ciccarello_collision_2017,Fischer, maffei_closed-system_2022}, $b(t,x)=\varrho^{-\frac{1}{2}}\sum_{k}e^{-i\omega_{k}(t-x/v)}a_{k}$ where $\varrho$ is the modes' density verifying the relation $\sum_{k}e^{-i\omega_{k}(t-t')}/\varrho=\delta(t-t')$. Finally $\gamma=g^{2}\varrho$ is the spontaneous emission rate of the 2LS. The operators $b(t,0)$ obey the bosonic algebra $\left[b(t,0),b^{\dagger}(s,0)\right]=\delta(t-s)$. In what follows and to lighten the notations, we will denote $b(t)=b(t,0)$.\\

We first consider a coherent input pulse of amplitude $\langle b(t)\rangle=\beta_t=(\beta/\sqrt{\varrho}) e^{-i\omega_0 t}$. It is convenient to solve the dynamics in the frame displaced by $\mathcal{D}(\beta)=\text{exp}\lbrace \int dt (b^{\dagger}(t)\beta_t-b(t)\beta^{*}_t)\rbrace$, such that the effective interaction term $\mathcal{D}(-\beta) H_{I}(t)\mathcal{D}(\beta)$ is the one of a resonant classical drive of Rabi frequency $\Omega/2=g\beta $ , and the input field is in the vacuum state. In the displaced frame, the joint light matter dynamics boils down to the one of a resonantly driven 2LS spontaneously emitting photons in the empty modes of the WG. If the 2LS is initially in its ground state, the joint state at time $\tau$ in the lab frame reads:
\begin{align}\label{eq_2LS}
    \ket{\Phi(\tau)}=\sqrt{P_{g}(\tau)}\ket{g,\phi_g(\tau)}+\sqrt{P_{e}(\tau)}\ket{e,\phi_e (\tau)}
\end{align}
with
\begin{align}\label{eq:Fwavefunction}
    &\ket{\phi_{\epsilon}(\tau)}=\frac{\mathcal{D}(\beta)}{\sqrt{P_{\epsilon}(\tau)}}\left[  \sqrt{p_{0,\epsilon}(\tau)} f^{(0,\epsilon)}(\tau) +\right.\\ \nonumber
&\left. +\sum_{n=1}^{\infty} \sqrt{p_{n,\epsilon}(\tau)} \int_0^{\tau} d\textbf{t}_n f^{(n,\epsilon)}(\tau;\textbf{t}_n)\prod_{i=1}^{n} b^{\dagger}(t_i)\right]\ket{0}.
\end{align}
 $\epsilon=g,e$ stands for ground and excited states of the 2LS, $\textbf{t}_n=\lbrace t_1,t_2,...t_n\rbrace$ with $t_1<t_2...<t_n<\tau$, $|f^{(0,\epsilon)}(\tau)|^2=1$ and $\int_0^{\tau} d\textbf{t}_n |f^{(n,\epsilon)}(\tau;\textbf{t}_n)|^2=1$ for any $n$. Finally $\sum_{n=0}^{\infty}p_{n,\epsilon}(\tau)=P_{\epsilon}(\tau)$. To save space, the explicit expressions of the functions $f^{(n,\epsilon)}(\tau;\textbf{t}_n)$ derived in~\cite{Fischer,maffei_closed-system_2022} are recalled in Appendix~\ref{app_1}. They reveal that in the long time limit $\gamma \tau\rightarrow \infty$, the input pulse has been scattered by the 2LS, which has relaxed in its ground state $\ket{g}$. The scattered light state is $\ket{\phi_g(\tau)}$, with $\gamma \tau \rightarrow \infty$. It involves the probabilities $p_{n,g}(\tau\rightarrow \infty) = p_{n}$. For $n\geq 1$, $p_{n}$ is the probability that the 2LS has scattered $n$ photons in other modes than the driving mode. Conversely the limit $p_{0} \rightarrow 1$ captures the case where the 2LS solely exchanges photons with the driving mode. This happens in the purely stimulated regime after a complete Rabi oscillation ($2\pi$ pulse), such that the 2LS is brought back in the ground state at the end of the interaction. It also happens in the so-called linear regime where the pulse Rabi frequency $\Omega$ is much lower than the spontaneous emission rate $\gamma$. Then the 2LS population remains vanishingly small all along the interaction with the pulse. In this case, the shape of the pulse remains almost unaltered, input and output pulses solely differing by a $\pi$ phase shift (See Appendix~\ref{app_1}).

When the input field is a single photon pulse $\ket{1}=\int dt \xi(t) b^{\dagger}(t)\ket{0}$ with $\int dt |\xi(t)|^2 =1$, and the initial state of the 2LS is the ground state, the joint state at time $\tau$ reads:
\begin{align}\label{eq:1p_solution}
\ket{\Phi(\tau)}&= \sqrt{\gamma}\tilde{\xi}(\tau) \ket{e,0}+\ket{g}\otimes\\ \nonumber
&\left( \int_{\tau}^{\infty} dt\xi(t) b^{\dagger}(t)+\int_{0}^{\tau} dt\varUpsilon(t)  b^{\dagger}(t)\right) 
\ket{0},
\end{align}
with $\tilde{\xi}(t)=e^{-\gamma t/2}\int_{0}^{t} dt' \left[e^{\frac{\gamma t'}{2}+i\omega_0 t'}\xi(t')\right]$ and $\varUpsilon(t)=\xi(t)-\gamma \tilde{\xi}(t)e^{-i\omega_{0} t}$, see Ref.~\cite{maffei_closed-system_2022} for the derivation. In the long time limit $\gamma \tau\rightarrow \infty$, the 2LS has decayed back in the ground state (Eq.~\eqref{eq:1p_solution}). The shape of a monochromatic input pulse is not altered by the scattering, input and output pulses solely differing by a $\pi$ phase shift~\cite{kojima_efficiencies_2004}.

\section{Measuring a spin with light}
We now focus back on the 4LS featuring the SPI. In analogy with the previous section, we treat the dynamics in the interaction picture with respect to the bare Hamiltonian $H_{\text{4LS}}+H_{f\text{pol}}$ (Eqs.~\eqref{eq_spinbare} and~\eqref{eq_field_bare}). The SPI coupling Hamiltonian (Eq.~\eqref{eq_spi_coupling}) in the interaction picture reads:
\begin{align}\label{eq_intpicspin}
H_{\text{SPI}}(t)=\hbar i \sqrt{\gamma}\sum_{j=\text{R,L}}\left[\sigma^{\dagger}_{j}(t)b_{j}(t)-b^{\dagger}_{j}(t)\sigma_{j}(t)\right],
\end{align}
where we defined the emitter's dipoles in the interaction picture as $\sigma_{j}(t)=e^{-i\omega_{0}t}\sigma_{j}$, and the annihilation operators destroying polarized excitations in $x=0$ at time $t$, $b_{j}(t)=\varrho^{-\frac{1}{2}}\sum_{k}e^{-i\omega_{k}t}a_{k,j}$. The operators $b_{j}(t)$ obey the bosonic algebra $\left[b_{j}(t),b_{j'}^{\dagger}(s)\right]=\delta_{j,j'}\delta(t-s)$ with $j,j'\in{\text{R,L}}$. Let us notice that, as we assumed the light-matter coupling to be the same for both polarizations, the two transitions of the 4LS have same decaying rate $\gamma$.

From the reminders on the 2LS, it appears that the spin states $\ket{\uparrow/\downarrow}$ are stable under the coupling with light in the long time limit $\gamma \tau \gg 1$. Indeed after the transitions $\ket{\up} \rightarrow \ket{\Uparrow}$ (resp. $\ket{\dwn} \rightarrow \ket{\Downarrow}$) have been driven and light pulses scattered, the spin is brought back to its initial state. In the limit of low-intensity, monochromatic pulses, the 2LS study also shows that the shape of the light pulses is unaltered by the interaction, input and output pulses solely differing by a $\pi$ phase shift.  This effect is essential to capture the physics at play in SPIs. For single, L-polarised (resp. R) photon pulses denoted $\ket{1_\text{L}}$ (resp. $\ket{1_\text{R}}$), it translates into the following map:
\begin{align}\label{eq:map_phase}
\ket{\up(\dwn),1_{\text{R(L)}}}&\rightarrow -\ket{\up(\dwn),1_{\text{R(L)}}},\\ \nonumber
    \ket{\dwn(\up),1_{\text{R(L)}}}&\rightarrow \ket{\dwn(\up),1_{\text{R(L)}}},
\end{align}
This map lies at the basis of several proposals for generating photonic gates~\cite{hu_giant_2008,bonato_cnot_2010} and more recently 2D photonic clusters~\cite{Pichler2017}. However, most functionalities of optical computing require operating at minimal speed, hence involve light pulses of finite duration. We now exploit our analytical model to explore the spin-light dynamics and its potential for quantum technologies beyond the monochromatic approximation. 

To benchmark the performances of the SPI, we choose to analyze it as a device, whose purpose is to perform QND measurements of the spin state in the $\ket{\uparrow/\downarrow}$ basis \cite{Grangier1998}. In this section we focus on the pre-measurement step \cite{neumann_mathematical_2018}: Starting from a well defined initial state, the light evolves conditionally to the spin state. While the spin state remains unaltered at the end of the process, the final light states $\ket{\psi_{\uparrow/\dwn}}$ become respectively correlated to $\ket{\uparrow/\downarrow}$: they are dubbed pointer states \cite{zurek_decoherence_2003}. Their overlap defines the performance of the pre-measurement and is quantified by the so-called quantum Bhattacharyya coefficient (qBhat)~\cite{fuchs_cryptographic_1999}:
\begin{align}\label{eq:qBattCoef}
\mathcal{B}_{q}=\vert\langle\psi_{\downarrow}\vert\psi_{\uparrow}\rangle\vert.
\end{align}
$\mathcal{B}_q = 0$ corresponds to orthogonal, hence perfectly distinguishable pointer states. It is this figure of merit we shall optimize throughout this section as a function of the light characteristics. In the spirit of quantum metrology \cite{maccone_science_2004}, we shall pay special attention to the light statistics in search for a quantum advantage. Namely, we shall compare the performance of the SPI as a quantum meter, depending if the probe is a coherent pulse that can be generated by a classical light source, or by a quantum pulse made of a coherent superposition of zero and one photon.
\\
\\
In the rest of this section we take as light initial state some horizontally (H) polarized pulse denoted $\ket{\psi_{\text{H}}}$. We first consider a coherent input pulse of amplitude $\alpha_t=\bra{\psi_{\text{H}}}b_{\text{H}}(t)\ket{\psi_{\text{H}}}$, with $b_{\text{H}}(t)\equiv\left(b_{\text{R}}(t)+b_{\text{L}}(t)\right)/\sqrt{2}$. The pointer states $\ket{\psi^\text{cs}_{\up/\dwn}}$ (the superscript cs refers to the coherent input state) can be found from the solution of the associated spinless problem Eqs.~\eqref{eq_2LS} and \eqref{eq:Fwavefunction} with $\beta_t=\alpha_t/\sqrt{2}$. They read:
\begin{align}\label{eq:CoherentFwavefunction}
\vert\psi_{\uparrow(\downarrow)}^{\text{cs}}\rangle&=\mathcal{D}^{(\text{H})}(\alpha)\left[ \sqrt{p_0} f^{(0)}+\right. \\ \nonumber
&+\left. \sum_{n=1}^{\infty} \sqrt{p_{n}}\int_{0}^{\infty} d\textbf{t}_n f^{(n)}(\textbf{t}_n)\prod_{i=1}^{n}b_{\text{R(L)}}^{\dagger}(t_i)\right]\ket{0}
\end{align}
where $f^{(0)}=\text{lim}_{\gamma \tau\rightarrow\infty}f^{(0)}(\tau)$ and $f^{(n)}(\textbf{t}_n)=\text{lim}_{\gamma \tau\rightarrow\infty}f^{(n,g)}(\tau;\textbf{t}_n)$, and $\mathcal{D}^{\text{(H)}}(\alpha)$~$=\exp\lbrace\int dt \left(b^{\dagger}_{\text{H}}(t)\alpha_t-b_{\text{H}}(t)\alpha^{*}_t\right)\rbrace$ is the displacement operator of the H-polarized propagating field's mode. $f^{(n)}(\textbf{t}_n)$ and $p_n$ do not depend on the light's polarization as the amplitude of the light-matter coupling is assumed to be the same for R and L, see Eq.~\eqref{eq:Int-pic-Hamiltonian}. Plugging Eq.~\eqref{eq:CoherentFwavefunction} into Eq.~\eqref{eq:qBattCoef} yields:
\begin{equation}
{\cal B}_{q}^{\text{cs}}=p_{0}.\label{eq:cf-qbhat}
\end{equation}
The physical meaning of Eq.~\eqref{eq:cf-qbhat} is transparent. Polarized photons scattered in the empty modes of the WG signal that the transition of the same polarization was driven, which carries information on the spin state.  Conversely in the limit where no photon was scattered ($p_0 \rightarrow 1$), the final light states are indistinguishable.

\begin{figure}
\includegraphics[width=0.5\textwidth]{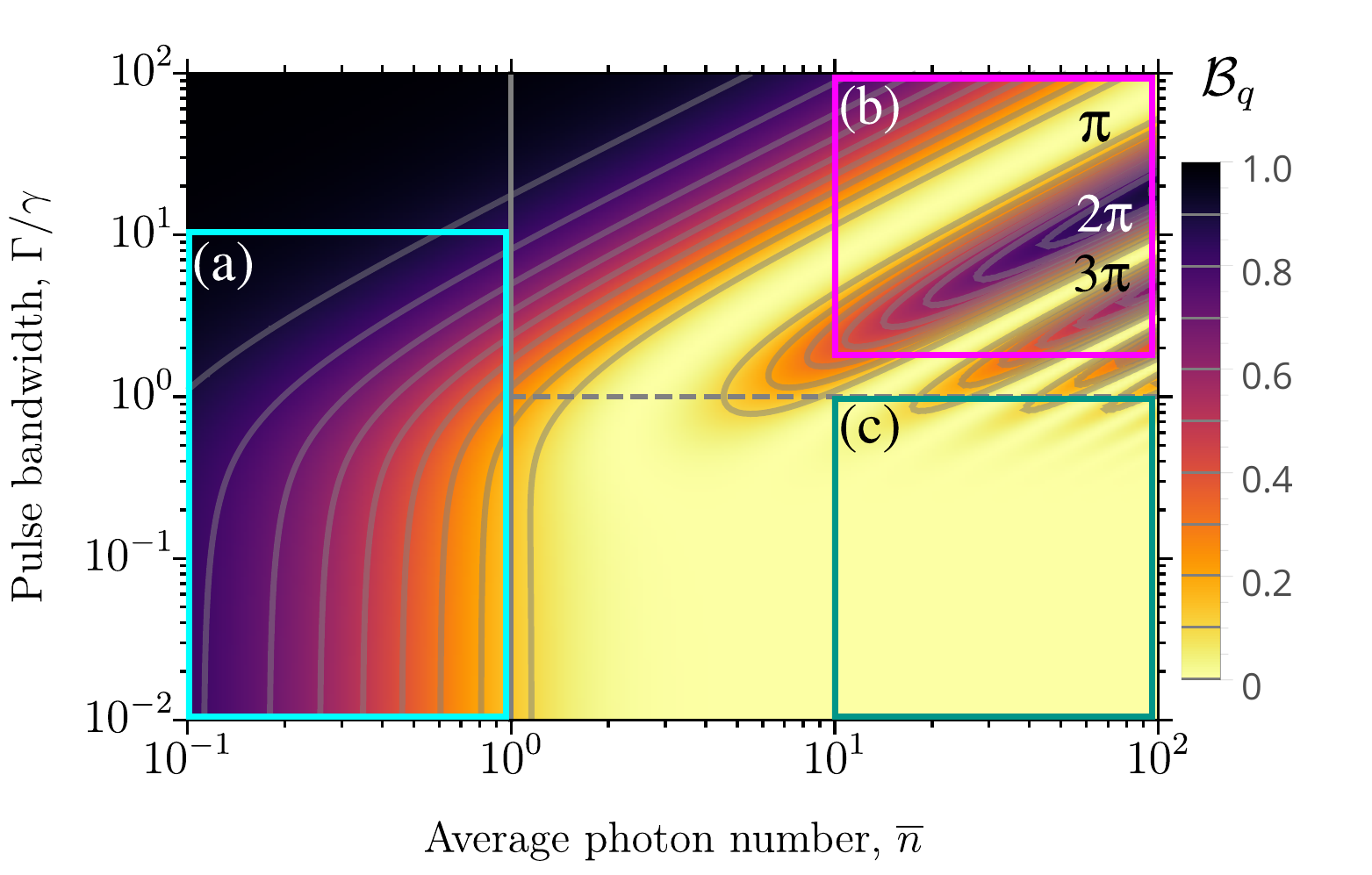}
\caption{Quantum Bhattacharyya coefficient for coherent, H-polarized input fields of amplitude $\alpha_t=\sqrt{\bar{n}\Gamma}e^{-\Gamma t/2-i\omega_0 t}$. The horizontal axis corresponds to the average number of photons per pulse, $\bar{n}$, and the vertical axis to the pulse bandwidth, $\Gamma$, in units of the decay rate $\gamma$. The color corresponds to the value of qBhat according to a color-scale going from yellow ($\mathcal{B}^{cs}_{q}=0$), to black ($\mathcal{B}^{cs}_{q}=1$). a) low-energy regime, b) high-energy, short-pulse regime where the system undergoes Rabi oscillations, and c) high-energy, long-pulses regime.}
\label{Fig2}
\end{figure}

Figure~\ref{Fig2} shows the value of qBhat for a coherent input pulse of amplitude $\alpha_t= \sqrt{\bar{n}\Gamma}e^{-\Gamma t/2-i\omega_0 t}$ as a function of the average photon number, $\bar{n}$, and the bandwidth, $\Gamma$. Regions (b) and (c) are the high-energy regions, $\bar{n}\geq 10$. The fringes in the region (b) capture Rabi oscillations~\cite{scully_quantum_1997}. Bright fringes correspond to complete inversions of the emitter's population ($\pi$ pulses), after which spontaneous emission occurs with certainty ($p_1=1$, $p_0=0$). When the spin is initially in a coherent superposition of $\ket{\up}$ and $\ket{\dwn}$, this situation leads to maximal spin-light entanglement, and was recently used to generate elementary 1D cluster states \cite{Fioretto2022,cogan2021deterministic} following the seminal proposal by Lindner and Rudolph~\cite{lindner_proposal_2009}. Conversely, dark fringes correspond to complete Rabi oscillations ($2\pi$ pulses), which leave the emitter in its ground state. Hence no photon is scattered in other modes than the driving mode, and no information can be extracted on the spin state. 

\begin{figure}
\includegraphics[width=0.5\textwidth]{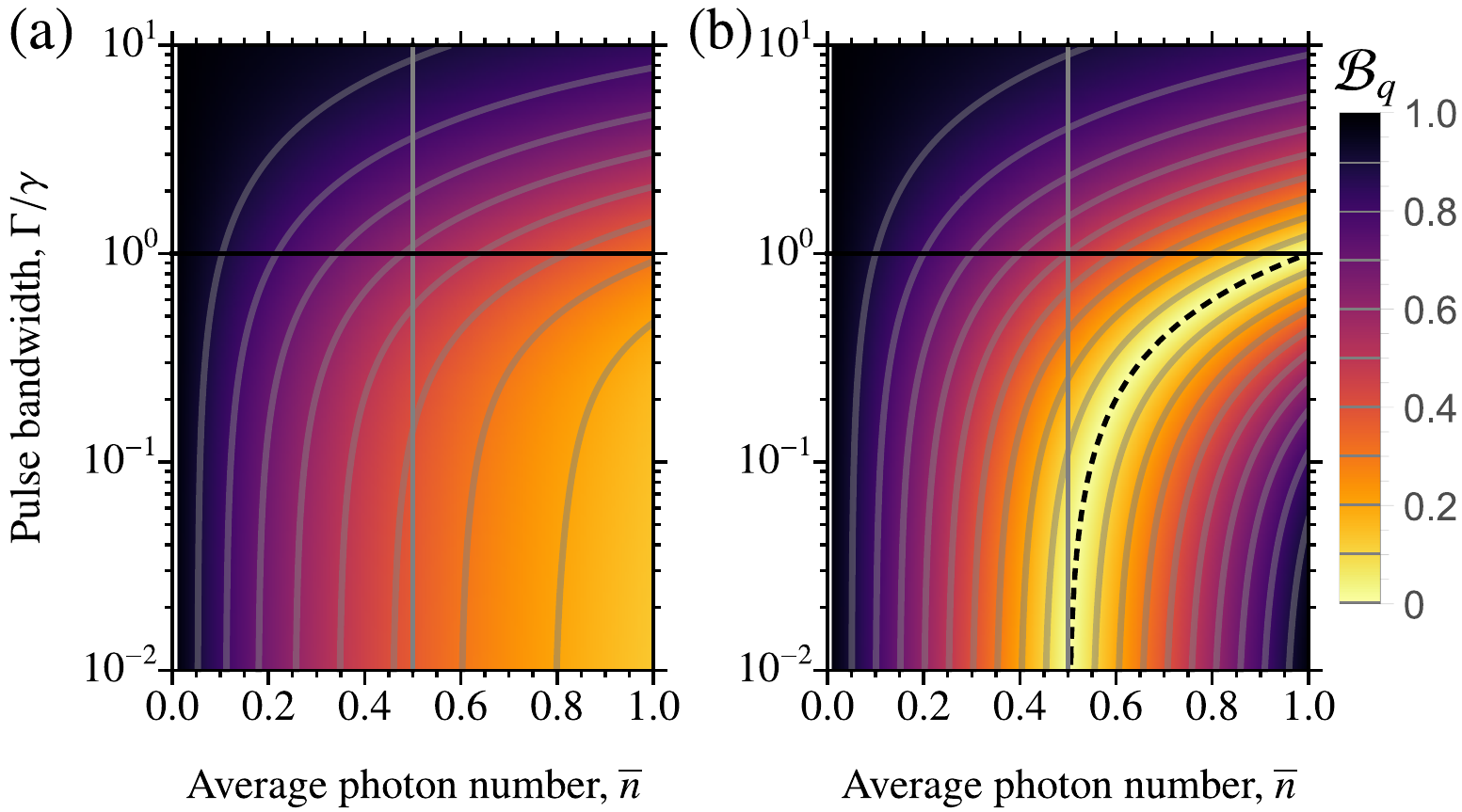}
\caption{Quantum Bhattacharyya coefficient in the low-energy regime. a) QBhat for coherent input fields. b) QBhat for superpositions of zero and single photon states. The fields are H-polarized and have same temporal profile being a decreasing exponential of rate $\Gamma$. The horizontal axes correspond to the average number of photons carried by the fields, and the vertical axes to their bandwidth, $\Gamma$, in units of the decay rate $\gamma$. The color corresponds to the value of qBhat, according to a color-scale going from yellow ($\mathcal{B}_{q}=0$), to black ($\mathcal{B}_{q}=1$). The gray line indicates $\bar{n}=0.5$, while the black line indicates the points where $\Gamma=\gamma$. In panel (a) the qBhat is always greater than zero, while in (b) it vanishes for all the points of the black dashed line of equation $\bar{n}=(1+\Gamma/\gamma)/2$.}
\label{Fig3}
\end{figure}

Region (a) ($\bar{n}\leq 1$) is the low-energy regime on which we focus from now on. As it appears on Fig.~\ref{Fig3}(a), $\mathcal{B}^{\text{cs}}_\text{q}$ never vanishes in this region, i.e. the pointer states never become perfectly distinguishable. Distinguishability fully vanishes in the monochromatic limit $\Gamma\ll\gamma$ (input pulses much longer than the lifetime of the atomic excited states) for a low energy input coherent field. This is the so-called linear regime mentioned in the previous section where the shapes of the scattered pulses remain almost unaltered, while undergoing a $\pi$ phase shift. Hence, the pointer states respectively read $\ket{\psi_{\up}} \approx \ket{\alpha/\sqrt{2}}_L\otimes \ket{-\alpha/\sqrt{2}}_R $  ($\ket{\psi_{\dwn}} \approx \ket{-\alpha/\sqrt{2}}_L\otimes \ket{\alpha/\sqrt{2}}_R $) - which strongly overlap for low energy pulses where $|\alpha|^2 \ll 1$.

We now take as input pulse a coherent superposition of zero and single photon states, $\ket{\psi_{\text{H}}}=c_{0}\ket{0}+c_{1}\vert1_{\text{H}}\rangle$, with $\vert c_{0}\vert^{2}+\vert c_{1}\vert^{2}=1$, $\vert1_{\text{H}}\rangle=\int dt \xi(t) b_{\text{H}}^{\dagger}(t)\ket{0}$. The pointer states $\ket{\psi^\text{qs}_{\up/\dwn}}$ (the superscript qs refers to the quantum statistics of this input state) can be found from the solution of the associated spinless problem Eq.~\eqref{eq:1p_solution} and read:
\begin{align}\label{eq:sol1P}
    \ket{\psi^{\text{qs}}_{\up(\dwn)}}&=\left[c_0 +\frac{c_1}{\sqrt{2}}\int_{0}^{\infty} dt\xi(t)b^{\dagger}_{\text{L(R)}}(t)\right.\\ \nonumber
    &\left.+\frac{c_1}{\sqrt{2}}\int_{0}^{\infty} dt\varUpsilon(t)b^{\dagger}_{\text{R(L)}}(t) \right]\ket{0}.
\end{align}
Plugging Eq.~\eqref{eq:sol1P} into Eq.~\eqref{eq:qBattCoef} yields the qBhat further denoted $\mathcal{B}^{\text{qs}}_\text{q}$. When the input field is an exponential wavepacket $\xi(t)=\sqrt{\Gamma}e^{-\Gamma t/2 -i\omega_0 t}$, the latter reads
\begin{equation}
{\cal B}_{q}^{\text{ns}}=1-\frac{2\bar{n}\gamma}{\gamma+\Gamma},\label{eq:ns-qbhat}
\end{equation}
meaning that perfectly distinguishable pointer states can be produced for any input energy $1/2\leq \bar{n}\leq 1$, provided that $\bar{n}\equiv|c_1|^2=(1+\Gamma/\gamma)/2$. $\mathcal{B}^{\text{qs}}_\text{q}$ is plotted in Fig.~\ref{Fig3}(b) as a function of $\Gamma/\gamma$ and $\bar{n}$. 

Let us first consider the limit of monochromatic input pulses ($\Gamma\ll\gamma$). The H-polarized photons read $\ket{1_{\text{H}}}= (\ket{1_{\text{L}}} + \ket{1_{\text{R}}})/\sqrt{2}$ and evolve under the map \eqref{eq:map_phase} into $i\ket{1_\text{V}}$ (resp. $-i\ket{1_\text{V}}$) if the spin is $\up$ (resp. $\dwn$). Thus, the pointer states read $\ket{\psi_{\up}} = (c_0\ket{0} + ic_1\ket{1_\text{V}})$ and $\ket{\psi_{\dwn}} = (c_0 \ket{0} - ic_1\ket{1_\text{V}})$. They are maximally distinguishable for $c_0=c_1$, for which $\bar{n}=1/2$. 

For shorter pulses, the spin-controlled phase shift respectively acquired by $\ket{1_{\text{L}}}$ and $ \ket{1_{\text{R}}}$ while being scattered induces the clockwise or counterclockwise rotation of the pulse polarization, together with an alteration of its shape (see Appendix~\ref{app2} for a detailed analysis). The final polarization and shape of each pointer state depend on the energy, shape and duration of the input pulse. Remarkably, mode-matched single photons characterized by $\Gamma \sim \gamma$ and $\bar{n}=1$ give rise to pointer states where information is mostly encoded in the polarization of $\ket{\psi_{\up}}$ $\left(\ket{\psi_{\dwn}}\right)$ that rotates from H to R (resp. L) during the interaction. In any case, the pulse shape is modified by the scattering process, which dilutes entanglement over polarization and temporal light degrees of freedom. Dealing with temporal degrees of freedom thus appears as a strong source of non-ideality for SPIs, both for information extraction at the classical level and quantum information protocols. This non-ideality cannot be avoided as photonic quantum computation requires photons of finite duration. 

The two studied situations reveal a quantum advantage for spin-light entanglement generation, better performances being reached by using quantum superpositions rather than coherent pulses. A similar effect has been observed in the context of machine learning \cite{milburn}. We elaborate on this quantum advantage in Section 6.

\section{Spin readout}

In measurement theory, pre-measurements are followed by collapses, where the quantum meter undergoes a projective measurement \cite{neumann_mathematical_2018}. Classical outcomes are expected to provide information on the measured quantum system.   A convenient figure of merit of the full measurement scheme is provided by the classical (cl) Bhattacharyya coefficient (cBhat)~\cite{fuchs_cryptographic_1999, andrewbook}. In the present case, the cBhat quantifies the overlap among the conditional probabilities  $p_{\up / \dwn}(x)$ of obtaining the classical outcome $x$ among the set ${\cal X}$, given that the spin is prepared in the state $\ket{\up/\dwn}$:   
\begin{equation}
B_{\text{cl}}=\sum_{x\in{\cal X}}\sqrt{p_{\up}(x)p_{\dwn}(x)}.\label{eq:ClassicalBhattCoef}
\end{equation}
$B_{\text{cl}}=1$ signals identical distributions where the measurement does not provide any information about the spin state; conversely $B_{\text{cl}}=0$ corresponds to disjoint distributions granting complete knowledge of the spin state. The cBhat is lower bounded by the qBhat (${\cal B}_{\text{q}}\leq B_{\text{cl}}$), which captures that correlations can be degraded while extracting information at the classical level. By optimizing the classical measurement scheme, equality can be reached when both Bhattacharyya coefficients vanish~\cite{fuchs_cryptographic_1999}, which means when the two pointer states are perfectly distinguishable and collapsed in the proper basis. 

\begin{figure}[!th]
\includegraphics[width=0.45\textwidth]{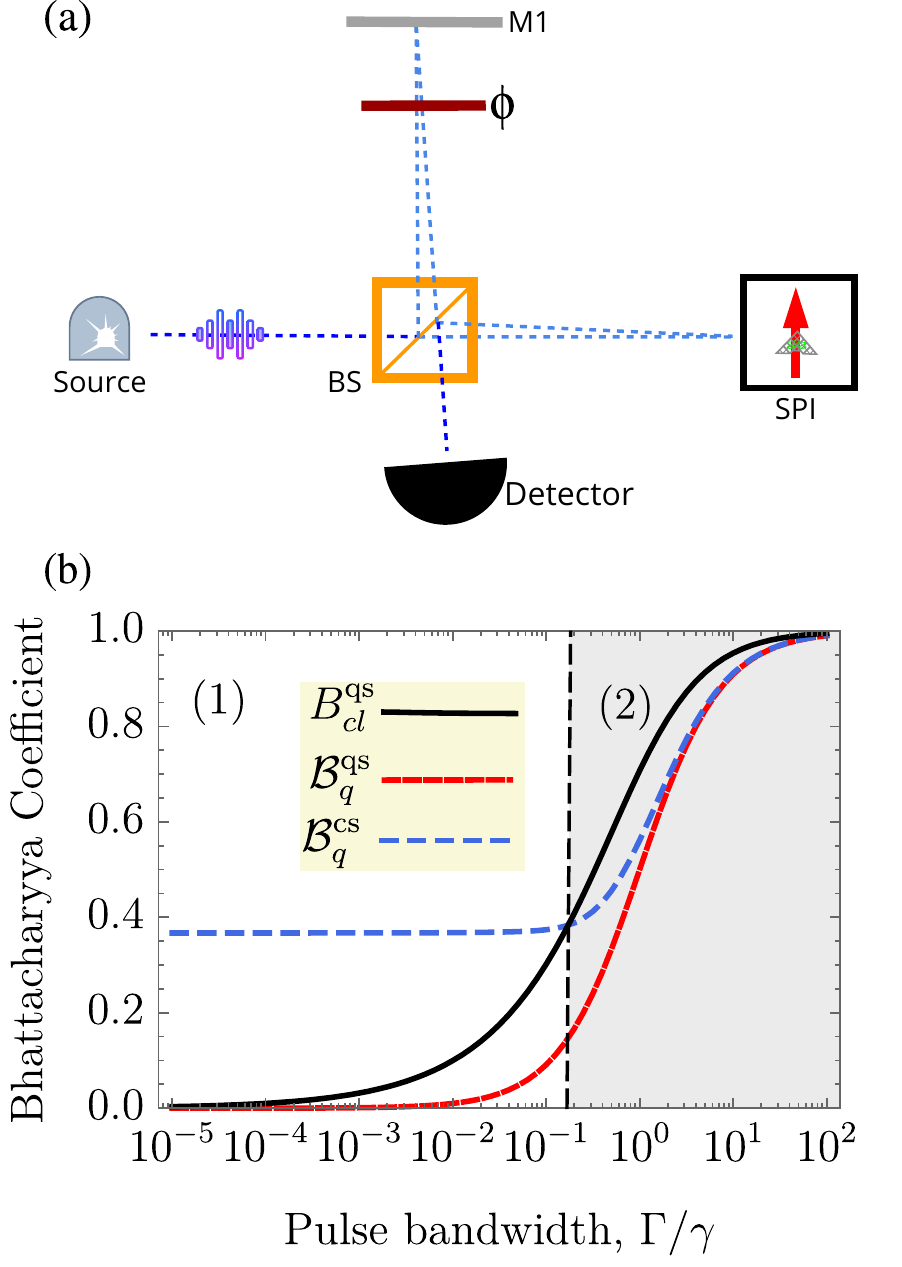}
\caption{Performance of the SPI for spin readout. a) Michelson interferometer. The right arm contains the SPI and the upper arm contains a tunable phase plate ($\phi$). BS is a 50/50 beam splitter. The source produces either a coherent pulse (classical state, cs) or a single photon (quantum state, qs) each with an average photon number of $\overline{n}=1$. b) Classical (black curve) and quantum (red dotted) Bhattacharyya coefficients for the quantum statistics, and quantum Bhattacharyya coefficient (blue dashed) for the classical statistics. All the coefficients are computed in the long-time limit as a function of pulse bandwidth $\Gamma$.}
\label{fig:readout}
\end{figure}

A possible measurement scheme including pre-measurement and collapse for a spin embedded in the SPI modeled above is depicted on Fig.~\ref{fig:readout}(a). The spin is initialized in $\ket{\up/\dwn}$ and then probed with a monochromatic R-polarized, low intensity pulse sent through a Michelson interferometer. As shown above in this regime, information on the spin state gets encoded in the phase of the pulse. The Michelson interferometer encompasses two balanced non-polarized beam splitters. One arm contains a tunable phase-plate ($\phi$), the other contains the SPI. The pulse exits the interferometer in one of the pointer states $\ket{\Psi_{\up}}$ and $\ket{\Psi_{\dwn}}$ that are respectively correlated with the spin up or down. This step is unitary and corresponds to the pre-measurement, whose performance is quantified by the qBhat, Eq.~\eqref{eq:qBattCoef}. The pulse is finally collapsed using a photo-counter positioned in the output port of the interferometer. The number of detected photons $x=0,1$ defines our two possible classical outcomes, whose statistics give rise to the cBhat assessing the performance of the full measurement scheme. In the following, the phase $\phi$ is chosen such that $p_{\up}(x)=1$ (resp. $0$) if the spin is $\ket{\up}$ (resp. $\ket{\dwn}$) for a monochromatic single photon input pulse. 

We analyze the performance of the readout if the spin is probed with a coherent or a single photon pulse, both characterized by an exponential temporal shape of spectral width $\Gamma$ and a mean number of photons $\bar{n}=1$. The overlap between the pointer states defines the two qBhat (quantum and coherent, resp. corresponding to the dotted red and the dashed blue curves). They are plotted on Fig.~\ref{fig:readout} as a function of the pulse bandwidth. Their behavior is consistent with Section 3: it reveals a quantum advantage for the pre-measurement step captured by ${\cal B}^{\text{cs}}_{\text{q}} \geq  {\cal B}^{\text{qs}}_{\text{q}}$, whichever the pulse bandwidth. The difference between the two qBhat, hence the quantum advantage is maximal in the monochromatic regime $\gamma \gg \Gamma$.  

We now focus on the robustness of the quantum advantage when the information on the spin state is extracted at the classical level. We have computed the evolution of the classical Batthacharyya coefficient in the case where the classical measurement is performed on a single photon pulse. The corresponding quantity is denoted $B_{cl}^{\text{qs}}$ and is plotted on Fig.\ref{fig:readout}(b) as a function of $\Gamma/\gamma$  (black solid curve). As expected, $B_{cl}^{\text{qs}}\geq \cal{B}_{\text{q}}^{\text{qs}}$, whichever the bandwidth, and the measurement scheme is optimal for monochromatic photons. Interestingly, the plot also reveals that sufficiently long pulses verify $B_{cl}^{\text{qs}}\leq \cal{B}_{\text{q}}^{\text{cs}}$, the latter inequality implies that $B_{cl}^{\text{qs}}\leq B_{cl}^{\text{cs}}$. In this regime (region $(1)$ on Fig.~\ref{fig:readout}), classical measurements performed on single photon pulses extract more information than on coherent fields of the same temporal profile and energy -- showing that the quantum advantage is robust and observable at the classical level. Note that it is not necessarily the case for shorter pulses where both the qBhat difference is lower and the classical measurement scheme is less adapted (region $(2)$ on Fig.~\ref{fig:readout}).

\section{Quantum advantage}
The analyses above point toward an energetic advantage of quantum nature when using quantum pulses as probes instead of coherent ones. Namely, quantum pulses show better performances for spin readout (as quantified by the quantum and classical Bhattacharyya coefficients) than classical ones with the same energy budget (as quantified by the mean number of photons per input pulse). Similar effects are observed in quantum metrology. It is interesting to compare the origin of such quantum advantages.

Firstly, let us recall that quantum metrology aims at maximizing the Fisher information about a parameter. The classical Fisher information is upper bounded by the quantum Fisher information, the inequality being saturated when an optimal measurement basis is chosen for the probe. This is a first similarity with the present situation which uses classical and quantum Battacharyya coefficients. Thus, the connecting factor between the two situations is information maximization: Fisher information in the case of quantum metrology, Battacharyya coefficient for the SPI. Note that the classical Battacharyya coefficient is equivalent to the Renyi information between the two distributions \cite{PhysRevLett.95.126803}, whereas the quantum Battacharyya coefficient is the overlap of the two pointer states, which is a measure of their distinguishability, also a critical concept in metrology \cite{jordan2015heisenberg}. Both cases give rise to a quantum advantage, because the use of quantum resources saturates the respective inequalities in order to obtain maximal information about either the parameter imprinted on the quantum state, or about which state the system of interest is in. 

Let us now examine the origin of the quantum advantage on concrete physical examples. In the case of metrology, we usually strive to measure the phase acquired by a probe as it interacts with a dispersive medium. The phase shift, hence the measurement precision is maximized with the variance of the probe Hamiltonian \cite{Wang2019}. For a fixed photon budget, this variance is larger for quantum superpositions ($N00N$ states) than for coherent states. Conversely in the SPI case under study, one wants to maximize the phase shift conditionally acquired by the probe as it interacts resonantly with the 4LS. This is a non-linear mechanism: the saturation of the quantum emitter appears as soon as the probe contains more than one photon, naturally altering the performance of coherent probes with respect to quantum superpositions of zero and one photons.  

Hence if the two mechanisms have different origins, they share the motivation of efficiency, in terms of budget per photon. Fine metrologic experiments rely on low intensity input pulses, giving rise to dedicated figures of merit such as the Fisher information per photon \cite{Pan2023}. In the same way, scaling up photonic quantum computers will require the optimization of the photonic resource cost \cite{QEI}: thus, our results are timely and show that quantum resources could play a significant role.

\section{Experimental feasibility} 
We finally discuss the experimental feasibility of SPIs as energy-efficient measuring devices of the spin state and their potential to show an energetic advantage of quantum nature. Let us first recall that in the past decade, degenerate SPIs as those presently studied have been implemented using dark excitons~\cite{schwartz_deterministic_2016} and more recently within atomic physics~\cite{thomas2022efficient,yang2021sequential} and semiconducting quantum dots in directional micropillar cavities~\cite{cogan2021deterministic,Fioretto2022}. The latter setting is nearly ideal, and was recently used to generate coherent superpositions of zero and one photon states of unprecedented purity~\cite{loredo_generation_2019}. Non-ideality stems from photon losses and phonon-induced dephasing~\cite{thomas2021race}, which we have included in our model to estimate the robustness of the quantum advantage (see Appendix~\ref{app3}).

We find that the quantum advantage region is reachable for setups having an overall efficiency no less than $80\%$, and a dephasing rate no more than $25\%$ of the optical decay rate $\gamma$. State-of-the-art quantum dots in directional cavities operating in the near IR can reach coupling efficiencies exceeding $90\%$ combined with low dephasing rates of 0.025$\gamma$ for $\gamma^{-1}\approx 100$~ps~\cite{tomm_bright_2021}, providing a promising direction towards experimental realization of the proposed setup. Commercially available superconducting nanowire detectors can also reach detection efficiencies exceeding $90\%$ in the near IR~\cite{Zhang_detectors}. In principle, the required quantum input light can be produced by a similar quantum dot device. However, this would compound the degrading effects of dephasing and would also bring the overall experimental efficiency below the $80\%$ bound. Near-term experimental realization could be possible using pulsed SPDC, with an appropriate bandwidth of at most $\Gamma=10^{-1}\gamma$. In this case, post-selecting on successfully-created single photons may bring the overall efficiency above the $80\%$ bound, allowing for an observation of the quantum advantage.\\
%
%
\section{Conclusion} We studied the interaction between a spin-carrying quantum emitter and a travelling pulse of light. We considered the low-energy regime where the light carries a maximum of one excitation in average, and compared a coherent field with a quantum superposition of zero and single photon states. We find that the latter state produces spin-light entanglement more efficiently than the former, providing an energetic quantum advantage. This quantum advantage is maintained when the information on the spin state is extracted by a classical agent who performs a projective measurement on the electromagnetic field after its interaction with the SPI. We showed that it can be observed within state-of-the-art physical implementations. Our study brings out a new interest in the exploitation of quantum resources based on energy efficiency, as already observed in quantum metrology. This inquiry is relevant from a fundamental point of view and useful to inspire new applications in the field of optical quantum computation, e.g. photon-photon gates and cluster states. Finally, our Hamiltonian model of SPI will be a valuable tool to mitigate errors in photonic quantum computations based on light pulses of finite duration.\\ 

\emph{Acknowledgments} We gratefully acknowledge financial support from the European Union Horizon 2020 Research and innovation Programme under the Marie Sklodowska-Curie Grant Agreement No. 861097, the Plan France 2030 through the project NISQ2LSQ ANR-22-PETQ-0006, the Foundational Questions Institute Fund (Grant No. FQXi-IAF19-01 and Grant No. FQXi-IAF19-05) , the John Templeton Foundation (Grant No. 61835), the ANR Research Collaborative Project “Qu-DICE” (Grant No. ANR-PRC-CES47), and the National Research Foundation, Singapore and A*STAR under its CQT Bridging Grant.


%
%

\pagebreak
\widetext
\newpage 

\setcounter{equation}{0}
\setcounter{figure}{0}
\setcounter{table}{0}
\setcounter{page}{1}
\renewcommand{\theequation}{A\arabic{equation}}
\renewcommand{\thefigure}{A\arabic{figure}}

\onecolumn
\appendix
\section{Explicit solution of the dynamics with coherent input field}\label{app_1}
Explicit expressions of the functions $f^{(n,\epsilon)}(\tau;\textbf{t}_n)$ in Eq.~\eqref{eq:Fwavefunction} can be derived with the method presented in Ref.~\cite{maffei_closed-system_2022}. When the initial state of the 2-level system is $\ket{g}$, the input pulse is a square pulse of $|\beta|^2$ photons and central frequency $\omega_0$, the functions $f^{(n,\epsilon)}(\tau;\textbf{t}_n)$ take the form:
\begin{align}
f^{(n,\epsilon)}(\tau;\textbf{t}_n)=\frac{F^{(n,\epsilon)}(\tau;\textbf{t}_n)}{\sqrt{p_{n,\epsilon}(\tau)}}.
\end{align}
Where
\begin{align}\label{eq:F0eg}
&F^{(0,g)}(\tau)=e^{-\gamma \tau/4}\left[\cos (\Omega' \tau/2)+\sin(\Omega' \tau/2)(\gamma)/(2\Omega') \right],\\ \nonumber
&F^{(0,e)}(\tau)=e^{-\gamma \tau/4}\sin(\Omega' \tau/2)\Omega/\Omega',
\end{align}
and for the terms with $n>0$:
\begin{align}
F^{(1,\epsilon)}(\tau;t_1)&=-\sqrt{\gamma} F^{(0,\epsilon)}(\tau-t_1) e^{-i\omega_0 t_1} F^{(0,e)}(t_1),\\ \nonumber
F^{(n>1,\epsilon)}(\tau;\textbf{t}_j)&=(-\sqrt{\gamma})^j F^{(0,\epsilon)}(\tau-t_{n}) e^{-i\omega_0 t_{n}}\left[ \prod_{i=2}^{n} F^{(0,e)}(t_{i}-t_{i-1})e^{-i\omega_0 t_{i-1}}\right] F^{(0,e)}(t_1),
\end{align}
with $\Omega=2\sqrt{\gamma/\varrho}\beta$, $\Omega'=\sqrt{(\Omega)^2-\gamma^2/4}$, and 
\begin{align}
p_{n,\epsilon}(\tau)=\int_0^{\tau} d\textbf{t}_n |F^{(n,\epsilon)}(\tau;\textbf{t}_n)|^2.
\end{align}

The average value of the operator $b(t)$ on Eq.~\eqref{eq:Fwavefunction} gives as expected the input-output relation:
\begin{align}\label{eq:in_out}
\langle b(t)\rangle = (\beta/\sqrt{\varrho}) e^{-i \omega_0 t} -\sqrt{\gamma}\langle \sigma(t)\rangle,
\end{align}
the term $\langle \sigma(t)\rangle$ can be computed analytically starting from the functions $F^{(n,\epsilon)}(t;\textbf{t}_n)$, it reads:
\begin{align}\label{eq:dipole}
\langle \sigma(t)\rangle = e^{-i\omega_0 t} F^{(0,e)}(t)(F^{(0,g)}(t))^*+e^{-i\omega_0 t} \sum_{n=1}^{\infty}\int_{0}^{t} d \textbf{t}_{n} F^{(n,e)}(t;\textbf{t}_{n})(F^{(n,g)}(t;\textbf{t}_{n}))^*
\end{align}
\subsection{Linear-regime: $\pi$ phase shift}
In the limit of very weak driving, i.e. $\Omega\ll \gamma$, the interaction with the atom leaves almost unchanged the shape of the field just producing a $\pi$-phase shift on the scattered field we hence have:
\begin{align}
\langle b(t)\rangle\approx-(\beta/\sqrt{\varrho})e^{-i\omega_0 t}, 
\end{align}
or equivalently from Eq.~\eqref{eq:in_out}:
\begin{align}\label{eq:linear_rel}
\langle \sigma(t)\rangle\approx(2/\sqrt{\gamma}) (\beta/\sqrt{\varrho})e^{-i\omega_0 t}.
\end{align}
The above equality is indeed verified when $\Omega\ll \gamma$ and hence $\Omega'=\sqrt{(\Omega)^2-\gamma^2/4} \approx i \gamma/2$.  Replacing $\Omega' \approx i \gamma/2$ in Eqs~\eqref{eq:F0eg} we find:
\begin{align}
F^{(0,g)}(t)&\approx 1;\\ \nonumber
F^{(0,e)}(t)&\approx(2/\sqrt{\gamma}) \left[ 1-e^{-\gamma t/2}\right](\beta/\sqrt{\varrho}).
\end{align}
Injecting the above expressions in Eq.~\eqref{eq:dipole} we find that Eq.~\eqref{eq:linear_rel} is verified as soon as $e^{-\gamma t/2}$ becomes negligible.

\section{Analysis of polarization and amplitude of the output field}\label{app2}
\subsection{Polarization}\label{app2_1}
The instantaneous polarization vector conditioned on the spin state is a 3-components vector, $\mathcal{E}^{\up(\dwn)}(t)=\lbrace \epsilon_z^{\up(\dwn)}(t), \epsilon_x^{\up(\dwn)}(t),\epsilon_y^{\up(\dwn)}(t)\rbrace$, defined as:
 \begin{align}
 \epsilon_z^{\up(\dwn)}(t)=\frac{I_{R}^{\up(\dwn)}(t)-I_{L}^{\up(\dwn)}(t)}{I^{\up(\dwn)}(t)};\\ \nonumber
 \epsilon_x^{\uparrow}(t)=\frac{I_{H}^{\up(\dwn)}(t)-I_{V}^{\up(\dwn)}(t)}{I^{\up(\dwn)}(t)};\\ \nonumber
 \epsilon_y^{\up(\dwn)}(t)=\frac{I_{A}^{\up(\dwn)}(t)-I_{D}^{\up(\dwn)}(t)}{I^{\up(\dwn)}(t)};\\ \nonumber
 \end{align}
 where
  \begin{align}
 I^{\up(\dwn)}_{j}(t)=\bra{\psi_{\up(\dwn)}(t)} b^{\dagger}_j(t)b_j(t)\ket{\psi_{\up(\dwn)}(t)}+\bra{\psi_{\U(\D)}(t)} b^{\dagger}_j(t)b_j(t)\ket{\psi_{\U(\D)}(t)},
 \end{align}
 and
 \begin{align}
b_{\text{H}}(t)=\frac{1}{\sqrt{2}}\left(b_{\text{R}}(t)+b_{\text{L}}(t)\right);\\ \nonumber
    b_{\text{V}}(t)=\frac{e^{i\pi/2}}{\sqrt{2}}\left(b_{\text{R}}(t)-b_{\text{L}}(t)\right);\\ \nonumber
    b_{\text{A}}(t)=\frac{e^{i\pi/4}}{\sqrt{2}}\left(b_{\text{R}}(t)-ib_{\text{L}}(t)\right);\\ \nonumber
      b_{\text{D}}(t)=\frac{e^{-i\pi/4}}{\sqrt{2}}\left(b_{\text{R}}(t)+ib_{\text{L}}(t)\right);\\ \nonumber
\end{align}
Since the input fields are H-polarized, the interaction excites symmetrically the states with opposite spin projections, i.e. $\langle \sigma_{\text{R}}(t)\rangle=\langle \sigma_{\text{L}}(t)\rangle$, we have:
    \begin{align}
        \epsilon_z^{\uparrow}(t)=- \epsilon_z^{\downarrow}(t);\\ \nonumber
        \epsilon_x^{\uparrow}(t)= \epsilon_x^{\downarrow}(t);\\ \nonumber
         \epsilon_y^{\uparrow}(t)=- \epsilon_y^{\downarrow}(t);
    \end{align}
    
\begin{figure}
    \centering
    \includegraphics[scale=0.3]{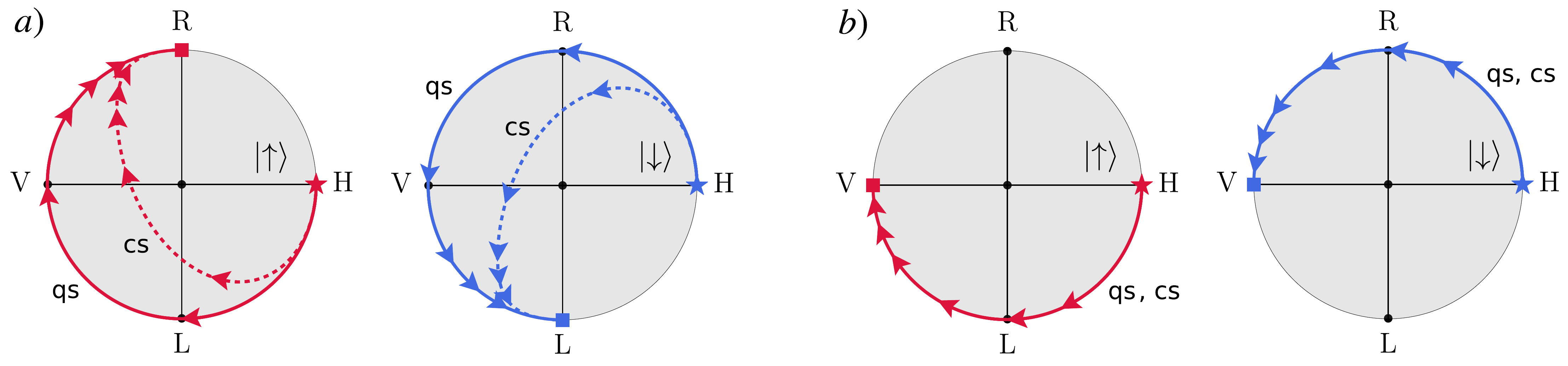}
    \caption{Instantaneous polarization vector trajectories through the R-H plane of the Poincar\'{e} sphere for different light-matter interaction conditions. a) Trajectories in the mode-matched regime where $\Gamma=\gamma$ for initial spin state $\ket{\uparrow}$ (red curves) or $\ket{\downarrow}$ (blue curves). The initial state of light has either classical statistics (cs) or quantum statistics (qs). b) Trajectories in the quasi-monochromatic regime where $\Gamma\ll\gamma$. Both qs and cs converge to the same polarization dynamics in this limit. The arrows on all plots indicate the time evolution over a timescale of ${\sim}10\gamma^{-1}$.}
    \label{sfig:polarization}
\end{figure}

    The instantaneous polarization vectors $\mathcal{E}^{\up(\dwn)}(t)$ can be equivalently defined by a couple of angles, the polar angle $\theta_{\up(\dwn)}(t)$, and  the azimuthal angle $\phi_{\up(\dwn)}(t)$. These angles can be obtained from the corresponding vector:
    \begin{align}
         \theta_{\up(\dwn)}(t)=\arccos{\left[\frac{\epsilon^{\up(\dwn)}_z(t)}{|\mathcal{E}^{\up(\dwn)}(t)|}\right]}\\ \nonumber
         \phi_{\up(\dwn)}(t)= \arctan{\left[\frac{\epsilon^{\up(\dwn)}_y(t)}{\epsilon^{\up(\dwn)}_{x}(t)}\right]}
      \end{align}
Now the polarization vector can be represented as a point on a sphere of radius 1, where the north and the south poles correspond respectively to the polarization state R and L. For both coherent and number statistics, we focus on two regimes with different energy and spectral bandwidth: (i) $\lbrace \bar{n}=1, \Gamma=\gamma \rbrace$ and (ii) $\lbrace \bar{n}=0.5, \Gamma=10^{-2}\gamma \rbrace $. In (i) the polarization vectors $\mathcal{E}_{\up}$ and $\mathcal{E}_{\dwn}$ rotate on the Poincar\'e sphere in opposite directions during the interaction reaching the orthogonal states, R and L, in the long-time stationary state (see Fig.~\ref{sfig:polarization}a). In (ii) the polarization vectors also rotates in opposite directions, but reach the same state, V, in a time shorter than the pulse duration, i.e. before the stationary state is reached (see Fig.~\ref{sfig:polarization}b).

\subsection{Amplitude}\label{app2_2}
For a fixed value of the instantaneous polarization, i.e. fixing the values of the angles $\theta_{\up(\dwn)}(t)$ and $\phi_{\up(\dwn)}(t)$ at time $t$, we can define the operator: 
    \begin{align}
     b_{\up(\dwn)}(t)=\cos{\left(\frac{\theta_{\up(\dwn)}(t)}{2}\right)} b_{\text{R}}(t)+\sin{\left(\frac{\theta_{\up(\dwn)}(t)}{2}\right)}e^{-i\phi_{\up(\dwn)}(t)}b_{\text{L}}(t).
      \end{align}
     The average value of the above operator give the average field's complex amplitude conditioned on the spin state:
\begin{align}
    \langle b_{\up}(t)\rangle=\cos{\left(\frac{\theta^{\up}(t)}{2}\right)}\left[\langle b^{in}_{\text{R}}(t)\rangle -\sqrt{\gamma}\langle \sigma_{\text{R}}(t)\rangle\right] + e^{-i\phi_{\up}(t)}\sin{\left(\frac{\theta^{\up}(t)}{2}\right)}\langle b^{in}_{\text{L}}(t)\rangle ;\\ \nonumber
     \langle b_{\dwn}(t)\rangle =e^{-i\phi_{\dwn}(t)}\sin{\left(\frac{\theta^{\dwn}(t)}{2}\right)}\left[\langle b^{in}_{\text{L}}(t)\rangle -\sqrt{\gamma}\langle \sigma_{\text{L}}(t)\rangle\right] +\cos{\left(\frac{\theta^{\dwn}(t)}{2}\right)}\langle b^{in}_{\text{R}}(t)\rangle,
     \end{align}
with $\langle b^{in}_{\text{R}}(t)\rangle=\langle b^{in}_{\text{L}}(t)\rangle=\langle b^{in}_{\text{H}}(t)\rangle/\sqrt{2}=\bra{\psi(0)}b_{\text{H}}(t)\ket{\psi(0)}/\sqrt{2}$ being the average amplitude of the input field prepared in the state $\ket{\psi(0)}$. 

We consider again the two different regimes of energy and spectral bandwidth (i) and (ii), and find the field's average amplitude in the long time limit. In the case (i), in the long time limit, the polarization vector $\mathcal{E}^{\up}$ reaches R and $\mathcal{E}^{\dwn}$ reaches L, i.e. $\phi^{\up}=\phi^{\dwn}=0$, $\theta^{\up}=0$ and $\theta^{\dwn}=\pi$. Plugging these values in the expressions above, we find that the average values of the field's amplitude conditioned on the spin states are identical in the long time limit. For this reason, in case (i), the polarization features a better pointer for the spin state than the field's mean amplitude. On the contrary in case (ii), both polarization vectors $\mathcal{E}^{\up}$ and $\mathcal{E}^{\dwn}$ reach V after a short transient of time, i.e. $\phi^{\up}=\phi^{\dwn}=0$, and $\theta^{\up}=\theta^{\dwn}=-\pi/2$. Plugging these values in the expression above, we find that the average values of the field's amplitude conditioned on the spin states have a relative phase of $\pi$. Then, in case (ii), the imaginary part of the field's amplitude (field's phase quadrature) features a better pointer than the polarization.
\section{Numerical methods}\label{app3}
To simulate the dynamics of measuring the spin state using a pulse of quantum light, we make use of the SLH framework \cite{combes_slh_2017} by considering a virtual source of the pulse \cite{kiilerich_input-output_2019}. This approach is based on the input-output formalism introduced by Gardiner and Collett \cite{gardiner_input_1985}, which was further developed by Gardiner \cite{gardiner_driving_1993} and Carmichael \cite{carmichael_quantum_1993}. Hence, it is a framework valid for the Markovian limit of light-matter interaction where there is no back-action on the source, and where there is dispersionless propagation of the field between components. Under these assumptions, the SLH framework provides solutions matching the analytical expressions given in the main text that have been obtained from the collision-model approach presented in~\cite{maffei_closed-system_2022}. 

The input pulse is modelled using degenerate two-mode virtual cavity source (s) whose evolution is fully described by an SLH triple $G_\mathrm{s}=(S_\mathrm{s},\textbf{L}_\mathrm{s},H_\mathrm{s})$, where the unitary map $S_\mathrm{s}=I$ is the identity operator, $\textbf{L}_\mathrm{s}=\sqrt{\Gamma}({a}_\mathrm{R},{a}_\mathrm{L})$ is the vector of collapse operators, ${H}_\mathrm{s}=\hbar\omega_\mathrm{s}{a}^\dagger_\mathrm{R}{a}_\mathrm{R}+\hbar\omega_\mathrm{s}{a}^\dagger_\mathrm{L}{a}_\mathrm{L}$ is the cavity Hamiltonian, ${a}_\mathrm{R}$ (${a}_\mathrm{L}$) is the right (left) circularly polarized cavity mode photon annihilation operator, $\Gamma$ is the cavity decay rate (pulse bandwidth), and $\hbar\omega_\mathrm{s}$ is the photon energy for both polarization modes. The spin-photon interface at zero magnetic field can be modelled as a 4-level atom composed of two degenerate 2-level transitions with the frequency $\omega_0$ and circular-polarized selection rules. The SLH triple for this system is similar to the source cavity: $G_0=({S}_{0},\textbf{L}_{0},{H}_{0})$, where ${S}_\mathrm{0}={I}$, $\textbf{L}_0=\sqrt{\gamma}({\sigma}_\text{R},{\sigma}_\text{L})$, ${H}_0=\hbar\omega_0{\sigma}^\dagger_\mathrm{R}{\sigma}_\mathrm{R}+\hbar\omega_0{\sigma}^\dagger_\mathrm{L}{\sigma}_\mathrm{L}$. 

 Using the series rule for SLH triples \cite{combes_slh_2017}, the cascaded system evolution is described by $G_0\triangleleft G_{p}=({I},\textbf{L},{H})$ where the collapse operators are $\textbf{L}=\textbf{L}_0+\textbf{L}_\mathrm{s}$ and the Hamiltonian is ${H}={H}_0+{H}_\mathrm{s}+{V}$ where the cascaded interaction potential is a sum of two Jaynes-Cummings coupling terms ${V}=-(i\sqrt{\Gamma\gamma}/2)[({\sigma}^\dagger_\text{R}{a}_\text{R}-{\sigma}_\text{R}{a}^\dagger_\text{R})+({\sigma}^\dagger_\text{L}{a}_\text{L}-{\sigma}_\text{L}{a}^\dagger_\text{L})]$. The cascaded system master equation is then $\dot{\rho} =\mathcal{L}{\rho}= -(i/\hbar)[{H},{\rho}]+\sum_k\mathcal{D}({L}_k){\rho}$, where the dissipator superoperator is $\mathcal{D}({L})=\mathcal{J}({L})-(1/2)\mathcal{A}({L})$,  $\mathcal{J}({L}){\rho}={L}{\rho}{L}^\dagger$ is the jump superoperator and $\mathcal{A}({L}){\rho}=\{{L}^\dagger{L},{\rho}\}$ is the anti-commutation (or amplitude damping) superoperator. Here, the Lindblad operator ${L}_k$ is the element of $\textbf{L}$ corresponding to polarization $k\in\{\text{R},\text{L}\}$. These operators also describe the total system input-output relations: ${a}_{k,\text{in}}-{a}_{k,\text{ out}}={L}_k=\sqrt{\gamma}{\sigma}_k+\sqrt{\Gamma}{a}_k$, where ${a}_{k,\text{in}}$ is the vacuum input mode to the cavity and ${a}_{k,\text{out}}$ is the output mode after the spin-photon interaction. The solution to the system dynamics is then formally given by ${\rho}(t)=\mathcal{K}(t,t_0){\rho}(t_0)$, where $\mathcal{K}(t,t_0) = \mathcal{T}e^{\int_{t_0}^t\mathcal{L}(t^\prime)dt^\prime}$, and $\mathcal{T}$ is the time-ordering operator. The system can then be solved with standard numerical integration techniques with an accuracy limited by the necessary truncation of the cavity energy levels. 
 
 This approach can account for pure dephasing and inefficiencies \cite{motzoi_continuous_2015,kiilerich_input-output_2019} with a few modifications to the master equation. To capture the pure dephasing rate of the light-matter interaction, we add the additional terms $\gamma^\star\mathcal{D}({\sigma}^\dagger_\text{R}{\sigma}_\text{R})+\gamma^\star\mathcal{D}({\sigma}^\dagger_\text{L}{\sigma}_\text{L})$. 
 To account for inefficiencies, we add a factor $\sqrt{\eta}$ on the virtual cavity mode operators: ${a}_k\rightarrow\sqrt{\eta}{a}_k$. This is also equivalent to reducing the cross-section of the light-matter interaction.

\subsection{QBhat}\label{app3_1}

For a pure initial cavity and spin state, the qBhat, $\mathcal{B}_{q}(t)=\vert\langle\psi_{\downarrow}(t)\vert\psi_{\uparrow}(t)\rangle\vert$, can be equivalently given by the normalized magnitude of spin coherence remaining at time $t$: $\mathcal{B}_q(t)=\left|\braket{{\sigma}_{\uparrow\downarrow}}\right|/c_\uparrow^*c_\downarrow$, where ${\sigma}_{\uparrow\downarrow}=\ket{\uparrow}\bra{\downarrow}$. We initialize the system in the state $\ket{\Psi(0)}=(\ket{\uparrow}+\ket{\downarrow})/\sqrt{2}\otimes\ket{\psi(0)}_{\text{s}}$, where s stands for source, so that $c_\uparrow=c_\downarrow=1/\sqrt{2}$. Hence, $\mathcal{B}_{q}(t)=2\left|\text{Tr}\left[{\sigma}_{\uparrow\downarrow}\mathcal{K}(t,0){\rho}(0)\right]\right|$, with ${\rho}(0)=\ket{\Psi(0)}\bra{\Psi(0)}$. To test the performance of the two different input pulse states, we either set $\ket{\psi(0)}_{\text{s}}=(\sqrt{1-\overline{n}}+\sqrt{\overline{n}}{a}^\dagger_\mathrm{H})\ket{0}$ for the number superposition pulse, or $\ket{\psi(0)}_{\text{s}}=e^{\alpha{a}_\mathrm{H}^\dagger-\alpha^*{a}_\mathrm{H}}\ket{0}$ for the coherent pulse, where $\alpha=\sqrt{\overline{n}}$, for $0\leq \overline{n}\leq 1$. The source produces a pulse of light from the quantum state prepared in the cavity with a temporal profile dictated by $\Gamma$. In the case that $\Gamma$ is constant, the pulse profile is a mono-exponential decay, which is the primary case studied in the main text. However, it is possible to shape the input pulse amplitude $f(t)$ by modulating $\Gamma$ in time using the formula $\Gamma(t) = |f(t)|^2/(1-\int_{0}^t|f(t^\prime)|^2dt^\prime)$. In Ref. \cite{kiilerich_input-output_2019}, they extend this approach to define a pulse with an arbitrary complex temporal wavefunction.

\subsection{Classical measurement}\label{app3_2}
In the main text we propose a classical measurement of the spin using a Michelson interferometer. The input field, being a single-photon field, passes through a balanced BS whose arms contain respectively the spin-photon interface, and a tunable phase shifter. Combining the map given by the spin-photon interaction, i.e. $\ket{\up(\dwn)}\ket{\psi_{\text{in}}}\rightarrow \ket{\up(\dwn)}\ket{\psi_{\up(\dwn)}}$, with that of the BS, i.e. $\ket{1,0}\rightarrow \left(\ket{1,0}+\ket{0,1}\right)/\sqrt{2}$ and $\ket{0,1}\rightarrow \left(\ket{0,1}-\ket{1,0}\right)/\sqrt{2}$, we can find the final state of the field, conditioned on the spin state:
\begin{align}
    \ket{\Psi_{\up(\dwn)}}= \frac{1}{2}\left(\ket{\psi_{\up(\dwn)}}\ket{0}-\ket{\psi_{\text{in}}}\ket{0}+\ket{0}\ket{\psi_{\up(\dwn)}}+\ket{0}\ket{\psi_{\text{in}}}\right).
\end{align}
The qBhat coefficient of this system, $\mathcal{B}^{qs}_{q}=|\bra{\Psi_{\up}}\Psi_{\dwn}\rangle|$ is identical to that computed considering the sole spin-photon interface interacting with an input field having $\bar{n}=0.5$.
In the considered lossless system, the presence or absence of light at the detector can be used to detect the spin's state. Then the cBhat can be written as $B_{cl}=\sqrt{p_\uparrow(\text{click}) p_\downarrow (\text{click})}+\sqrt{(1-p_\uparrow (\text{click}))(1-p_\downarrow (\text{click}))}$ where $p_j (\text{click})$ is the probability of a detection occurring when the spin is prepared in state $\ket{j}$, where $j\in\{\uparrow,\downarrow\}$.   

To numerically compute the cBhat, we split the input pulse with a balanced beam splitter, resulting in the modified SLH triple $G$ with $\textbf{L}=\textbf{L}_0+\textbf{L}_\mathrm{s}/\sqrt{2}$ and ${H}={H}_0+{H}_\mathrm{s}+{V}/\sqrt{2}$. 
The four output modes leaving the beam splitter after the interference are then given by $\left(\textbf{L}_\text{1},\textbf{L}_\text{2}\right)={U}_\text{BS}\left(\textbf{L},\textbf{L}_\mathrm{s}/\sqrt{2}\right)$, where ${U}_\text{BS}$ is a balanced beam splitter transformation operating on the interfering polarization modes within each vector operator.

The photon annihilation operator of the Michelson interferometer output mode monitored by the detector is given by $d_{\textbf{p}}=\textbf{p}\cdot\textbf{L}_1$, where $\textbf{p}$ is the polarization vector of the light. The probability $1-p_j$ of not measuring a photon in mode $d_{\textbf{p}}$ after one input pulse is given by $1-p_j=\lim_{t\rightarrow\infty}\text{Tr}\left[\rho_0(t)\right]$ \cite{wein_analyzing_2020} where the conditional state $\rho_0$ is found by removing the stochastic jump dynamics $\mathcal{J}(d_{\textbf{p}})$ induced by the quantum fluctuations of mode $d_{\textbf{p}}$ from the total cascaded system master equation: $\dot{\rho}_0(t)=\left(\mathcal{L}-\mathcal{J}(\hat{d}_{\textbf{p}})\right)\rho_0(t)$. In the proposed Michelson experiment, the measurement is based on the presence or absence of light at the detector, regardless of the polarization. Thus, the state ${\rho}_0$ conditioned on no detection is given by the equation of motion $\dot{\rho}_0=\left(\mathcal{L}-\mathcal{J}({d}_\text{R})-\mathcal{J}({d}_\text{L})\right){\rho}_0$, where $\textbf{L}_1=\left({d}_\text{R},{d}_\text{L}\right)$ as written in the circular polarization basis at the detector. Note that the equation of motion is unchanged if we apply a unitary transformation to $\textbf{L}_1$, meaning that the evolution of ${\rho}_0$ is independent of the polarization of light detected, as is expected.

\begin{figure}
    \centering
    \includegraphics[scale=0.6]{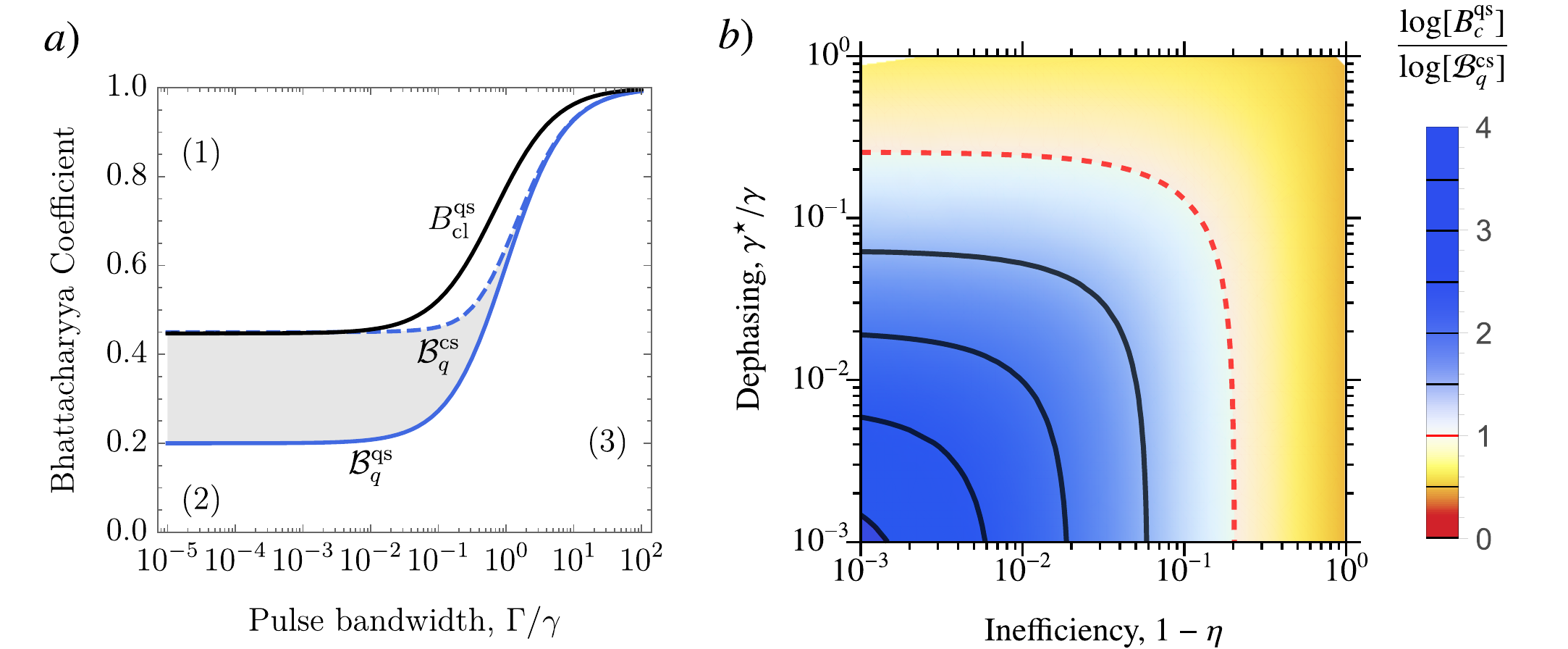}
    \caption{Bounds on the regime where the quantum advantage can be observed. a) Classical Bhattacharyya coefficient (black curve) for the Michelson interferometer scheme using a single photon input with an efficiency of $\eta=0.8$ and a trion with no dephasing $\gamma^\star=0$. The gray region between the quantum Bhattacharyya coefficients for the classical state input ($\mathcal{B}_q^\text{cs}$, dashed blue curve) and the quantum state input ($\mathcal{B}_q^\text{qs}$, solid blue curve) represents the region of quantum advantage for this scheme. The figure for $\eta=1$ and $\gamma^\star=0.25\gamma$ looks nearly identical. b) A map of the observed measurement efficiency relative to the classical bound in the monochromatic limit $\Gamma\ll\gamma$. The blue area below the red dashed curve shows the parameter region where the Michelson interferometer scheme using a single-photon input performs a better spin measurement than a coherent pulse of equal energy and shape.}
    \label{figsup:limits}
\end{figure}

Following the above approach, we compute the quantum and classical Bhattacharyya coefficients for a coherent state and a single photon input to the interferometer. When assuming ideal parameters $\eta=1$ and $\gamma^\star=0$, we acquire the plot presented in the main text Fig.~\ref{fig:readout}~b. By increasing $\gamma^\star$ while keeping $\eta=1$, we find that the cBhatt for the single photon input $B_{cl}^\text{qs}$ can be less than the qBhatt for the coherent state input $\mathcal{B}_q^\text{cs}$ only when $\gamma^\star/\gamma$ is less than around $0.25$, and this occurs in the monochromatic limit ($\Gamma\ll \gamma$). Similarly, by decreasing $\eta$ while keeping $\gamma^\star=0$, we find that $B_{cl}^\text{qs}<\mathcal{B}_q^\text{cs}$ only when $\eta$ is larger than about $0.80$, and this again occurs in the monochromatic limit (as illustrated in Fig. \ref{figsup:limits}~a). More generally, we look at the relative information extraction efficiency of the measurement defined by the ratio of the logarithms $\log(B_{cl}^\text{qs})/\log(\mathcal{B}_q^\text{cs})$. When this quantity exceeds 1, it implies that the Michelson interferometer implementation of the spin measurement using a single-photon pulse outperforms a coherent pulse of the same shape and input energy. It also implies that the Michelson scheme using a single photon outperforms any possible measurement scheme that uses a coherent pulse where an average photon number of $\overline{n}=1/2$ interacts with the spin system (see Fig. \ref{figsup:limits}~b).

\printbibliography

\end{document}